\documentclass[conference,a4paper]{IEEEtran}
\IEEEoverridecommandlockouts

\usepackage{cite}
\usepackage{amsmath,amssymb,amsfonts}
\usepackage{algorithmic}
\usepackage{graphicx}
\usepackage{textcomp}
\usepackage{xcolor}
\usepackage{bbm}
\def\BibTeX{{\rm B\kern-.05em{\sc i\kern-.025em b}\kern-.08em
    T\kern-.1667em\lower.7ex\hbox{E}\kern-.125emX}}

\begin{document}
\setlength{\abovedisplayskip}{2pt}
\setlength{\belowdisplayskip}{2pt}
\setlength{\abovedisplayshortskip}{2pt}
\setlength{\belowdisplayshortskip}{2pt}

\title{Sound Safeguarding for Acoustic Measurement Using Any Sounds: Tools and Applications
\thanks{KAKENHI JP23K20440 and JP21K19794 by JSPS (Japan Society for the Promotion of Science) supported this research.}
}

\author{\IEEEauthorblockN{Hideki Kawahara}
\IEEEauthorblockA{\textit{Center Innovation and Joint Research} \\
\textit{Wakayama University}\\
Wakayama, Japan \\
kawahara@wakayama-u.ac.jp}
\and
\IEEEauthorblockN{Kohei Yatabe}
\IEEEauthorblockA{\textit{Dept. Electrical Engineering and Computer Science} \\
\textit{Tokyo University of Agriculture and Technology}\\
Tokyo, Japan \\
yatabe@go.tuat.ac.jp}
\and
\IEEEauthorblockN{Ken-Ichi Sakakibara}
\IEEEauthorblockA{\textit{Dept. Speech-Language-Hearing Therapy} \\
\textit{Health Sciences University of Hokkaido}\\
Hokkaido, Japan \\
kis@hoku-iryo-u.ac.jp}
\and
\IEEEauthorblockN{Mitsunori Mizumachi}
\IEEEauthorblockA{\textit{Dept. Electrical and Electronic Eng.} \\
\textit{Kyushu Institute of Technology}\\
Kita-Kyushu, Japan \\
mizumach@ecs.kyutech.ac.jp}
\and
\IEEEauthorblockN{Tatsuya Kitamura}
\IEEEauthorblockA{\textit{Dept. Intelligence and Informatics} \\
\textit{Konan University}\\
Kobe, Japan \\
t-kitamu@konan-u.ac.jp}

}

\maketitle

\begin{abstract}
We demonstrate tools and applications developed based on the method of ``sound safeguarding,'' which enables any sound to be used for acoustic measurements.
We developed tools for preparation, interactive and real-time measurement, and report generation.
We extended and modified the method during its development based on its application in various practical situations.
We have open-sourced these tools and encourage prospective users to use them to improve their acoustic environments.
\end{abstract}

\begin{IEEEkeywords}
impulse response, discrete Fourier transform, linear time invariant system, signal dependent distortion, disturbance
\end{IEEEkeywords}

\section{Introduction}
We proposed the method of ``sound safeguarding,'' which enables any sound to be applied to acoustic measurements~\cite{Kawahara2022asjSG}.
Popular acoustic test signals, such as chirp tones and pseudo-random noise~\cite{Stan2002aesJ}, sound unpleasant.
The unpleasant nature of these test signals prevents live monitoring of acoustic conditions when an audience is present.
Sound safeguarding solves this problem by making the presented content itself usable for acoustic measurements.
In this demonstration, we present tools and applications~\cite{Kawahara2023apsipa,kawahara2024cocosda} that extend and modify the original method based on tests in several practical situations.

\section{Background}
%

We proposed a systematic framework for acoustic measurement called RHAPSODEE (Refinement by post Hoc Analysis using Periodic Stimulation with Orthogonal Decomposition for Extra Exploration)~\cite{Kawahara2023apsipa,kawahara2024cocosda,kawaharaGitRAPHSODEE} that uses the unitary transform nature of the discrete Fourier transform (DFT) and its efficient implementation~\cite{Frigo1998icassp,yatabe2021jasj6}. We implemented interactive and real-time acoustic measurement tools based on RHAPSODEE, which consist of safeguarded signal-based measurement. We also conducted offline measurements using the safeguarded signals. We redesigned and developed tools and procedures based on the findings and issues from these trials. The significant differences from the original proposal are frequency-dependent thresholding and (depending on applications) the removal of the repetition of test signals.

The organization of this paper is as follows. First, we begin with the impulse response measurement based on the DFT, and then introduce the safeguarding principle and its recent extensions. Finally, we introduce revised tools and procedures for several practical application cases.

\section{Principles and issues}

Let $\mathbf{F}$ represents the DFT matrix consisting of $(p, q)$-th element $F_{p,q}=(1/\sqrt{L})\mathrm{e}^{\gamma (p-1)(q-1)}$ ($\gamma = -2\pi \sqrt{-1}/L$).  
Let the DFT of an $L$-dimensional vector $\mathbf{x}$ consisting of a discrete time signal represent an $L$-dimensional vector $\mathbb{X}$.
Then, the following equations define DFT and the inverse DFT. %
\begin{align}
\mathbb{X} = \mathbf{F}\mathbf{x}, \ \ \ \mathbf{x} = \mathbf{F}^{\mathrm{H}}\mathbb{X} ,
\end{align}
where the symbol $\mathrm{H}$ represents  the conjugate transpose.

\subsection{Impulse response}
The output $\mathbf{y}$ of a linear time invariant system (let its impulse response in the discrete time domain $\mathbf{h}$ and $\mathbb{H}$ in the frequency domain) for the input $\mathbf{x}$ is below:
\begin{align}
\mathbf{y} & = \mathbf{F}^{\mathrm{H}} \mathbb{Y} , \  \mbox{\textit{i.e.}}\ \ 
 \mathbf{y} = \mathbf{F}^{\mathrm{H}} \left(\mathbb{H} \odot \mathbb{X}  \right), 
\end{align}
where $\odot$ represents the Hadamard product.

Then, the inverse DFT of the element-wise division of the output  $\mathbf{Y}$ by the input $\mathbf{X}$ yields the impulse response  $\mathbf{h}$.
\begin{align}
\mathbf{h} & =  \mathbf{F}^{\mathrm{H}} \left(\mathbb{Y} \oslash \mathbb{X} \right), 
\end{align}
where $\oslash$ represents the Hadamard division.

\subsection{Signal safeguarding}
Acoustic measurements in the real world inevitably consist of the observation noise $\mathbf{r}$ and $\mathbb{R}$ (in the time and frequency).
The estimated impulse response $\mathbf{h_\mathrm{est}}$ using the observed singal $\mathbf{s}=\mathbf{y}+\mathbf{r}$ ($\mathbb{S}=\mathbb{Y}+\mathbb{R}$ in the frequency domain.) is below:
\begin{align}
\mathbf{h_\mathrm{est}}   & = \mathbf{F}^{\mathrm{H}} \!\left(\mathbb{S} \oslash \mathbb{X} \right) 
 = \mathbf{F}^{\mathrm{H}} \!\left[ \left(\mathbb{Y} \oslash \mathbb{X} \right) 
  +  \left(\mathbb{R} \oslash \mathbb{X} \right) \right] , \label{eq:estImp}
\end{align}
where the second argument in (\ref{eq:estImp}) represents the error due to observation noise.
It shows that small absolute-valued elements of $\mathbb{X}$ result in huge error.
This huge error is why arbitrary signals are generally irrelevant for acoustic measurements.

Signal safeguarding sets lower limits $\mathbb{T} = T\cdot \mathbf{1}$ in the frequency domain and makes all elements have larger absolute values than $T$.
The following equations show the procedure.
\begin{align}
 \mathbf{x}_{\mathrm{SG}} & = \mathbf{F}^{\mathrm{H}} \!\left[\mathbb{X}\! + \! K_{\mathbb{T}}(\mathbb{X}) \! \odot \! (\mathbb{T} \! -\! |\mathbb{X}|)\! \odot\! (\mathbb{X}  \oslash |\mathbb{X}|) \right] \nonumber \\
& = \mathbf{x}\! +\! \mathbf{F}^{\mathrm{H}} \!\left[K_{\mathbb{T}}(\mathbb{X}) \! \odot \! (\mathbb{T} \! -\! |\mathbb{X}|)\! \odot\! (\mathbb{X}  \oslash |\mathbb{X}|) \right] \label{eq:sgFreq}
\end{align}
where $K_{\mathbb{T}}(\mathbb{X})$ represents a binary (0, 1) valued vector function.
The $m$-th element of $K_{\mathbb{T}}(\mathbb{X})$ is 1 when $T<|X[m]|$, and is 0 otherwise.
Note $|\mathbb{X}|$ represents a vector consisting of the absolute value of each element of $\mathbb{X}$.
The equation (\ref{eq:sgFreq}) provides an interpretation that the signal safeguarding procedure adds (virtually stable and slight) noise to the original signal.

We extended the method by introducing frequency-dependent lower limits $T[m]$ and trailing zero-padding.
The latter enabled discarding repetition, if necessary, without losing the benefit of DFT-based convolution, namely, no need for sub-sampling associated with the spectrogram representation by short-term Fourier transform~\cite{oppenheim2024ieee}.

\section{Tools and Applications}
We introduced tools and procedures for each stage of acoustic measurements.
They are preparation, measurement, and report generation.

\begin{figure}[tbp]
\begin{center}
\includegraphics[bb = 0 0 247 99, width=0.9\hsize]{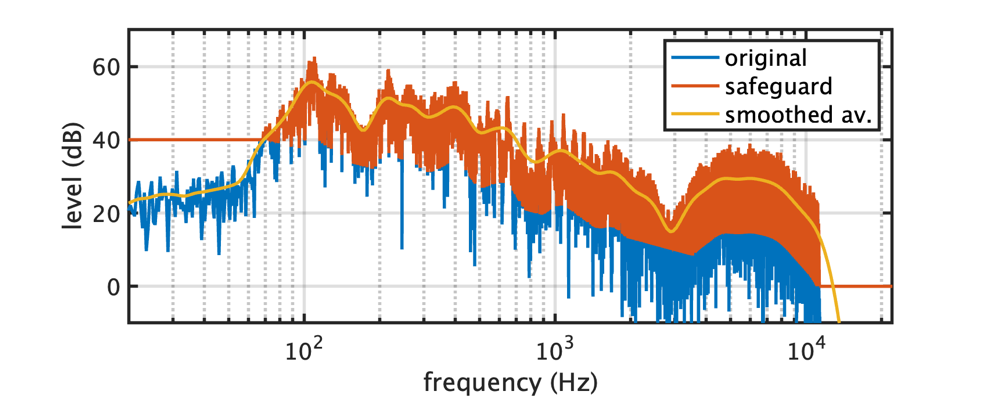}\\
\vspace{-3mm}
\caption{Power spectra of the original and the safeguarded test signal with smoothed power spectra and frequency-dependent threshold $\Theta[m]$. Sentence: ``This is a test signal for acoustic environment.''\label{fig:sgSample}}
\vspace{-6mm}
\end{center}
\end{figure}

\subsection{Stage 1: Preparation}
We implemented an interactive tool that converts any sound file into safeguarded test signals.
Figure~\ref{fig:sgSample} shows an example of the original and the safeguarded test signal.

\subsection{Stage 2: Measurement}
\begin{figure}[tbp]
\begin{center}
\includegraphics[bb = 0 0 247 176, width= 0.85\hsize]{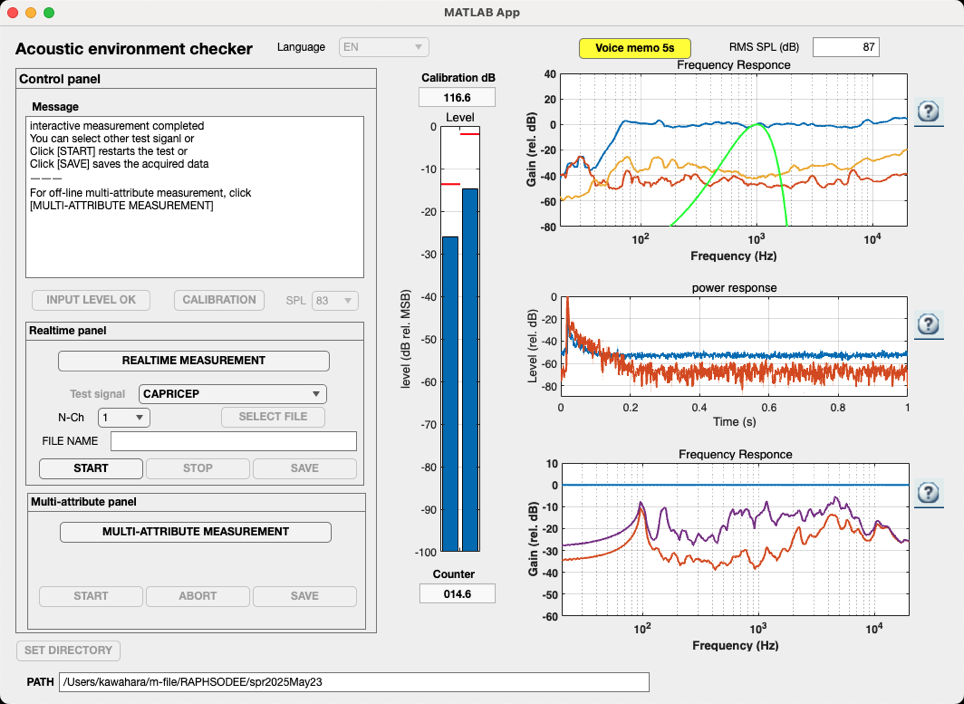}\\
\vspace{-3mm}
\caption{Snapshot of the GUI of the interactive tool.\label{fig:interavctivrTool}}
\end{center}
\vspace{-6mm}
\end{figure}

We revised an interactive test tool and added the ability to use any file for acoustic measurement.
It demonstrates the effectiveness of signal safeguarding by comparing the results using the original and the safeguarded files.
The tool saves the acquired system outputs, associating them with metadata about the safeguarded signal used.
We also used the safeguarded signals for offline measurements.
Figure~\ref{fig:interavctivrTool}shows a snapshot of the interactive tool.
The left side has the control buttons and message display.
The right-side graphs are a real-time monitor of the acoustic environment.
Refer to~\cite{Kawahara2023apsipa,kawahara2024cocosda} for details.

\subsection{Stage 3: Report}
All GUI operations, calibration data, and acquired signals are time-stamped and logged.
We are integrating analysis and report generation tools for the saved test results and logs.

\section{Conclusion}
We demonstrated tools and applications of ``signal safeguarding,'' which enables any sound to be applied to acoustic measurements.
We found that demonstrations using these tools are practical and valuable for students, helping them acquire a basic understanding and skills of sound acquisition and presentation.
We open-sourced the tools and related materials~\cite{kawaharaGitRAPHSODEE}.

\section*{Acknowledgment}

We appreciate  Drs. Masashi Unoki, Toshio Irino, Toshie Matsui, Eriko Aiba, and their students for preliminary acoustic measurement tests and comments.


\end{document}